# Default-all is dangerous!

Wolfgang Gatterbauer Alexandra Meliou Dan Suciu
*Department of Computer Science and Engineering,*
*University of Washington, Seattle, WA, USA*
{gatter,ameli,suciu}@cs.washington.edu

## Abstract

We show that the *default-all propagation* scheme for database annotations is dangerous. Dangerous here means that it can propagate annotations to the query output which are *semantically irrelevant* to the query the user asked. This is the result of considering all relationally equivalent queries and returning the union of their where-provenance in an attempt to define a propagation scheme that is insensitive to query rewriting.

We propose an alternative query-rewrite-insensitive (QRI) where-provenance called *minimal propagation*. It is analogous to the minimal witness basis for why-provenance, straight-forward to evaluate, and returns *all relevant and only relevant* annotations.

## 1 *Query-Rewrite-Insensitive* provenance

Provenance is sensitive to query rewriting unless carefully defined. Sensitive here means that the returned provenance may be different for a relationally equivalent query (we focus exclusively on conjunctive queries under set semantics). This is surprising at first since we are accustomed to leaving it to the database engine to choose the simplest relationally equivalent query to return our results. If we also consider provenance, then we are not guaranteed to get the provenance output we expect.

With this argumentation, Buneman et al. [2] proposed that it is important to find a clean semantics for provenance that guarantees to give the same result for relationally equivalent queries. At least two well-known query-rewrite-insensitive (QRI) versions have been defined: Buneman et al. [2] defined the *minimal witness basis* for why-provenance, and Bhagwat et al. [1] defined the *default-all propagation* scheme for where-provenance.

Our goal with this paper is to point to some semantic problems with the way the QRI property is achieved by default-all propagation. We also show how to fix these problems with an alternative propagation scheme.

| | Why-provenance | Where-provenance |
|---|---|---|
| Naive | witness [2] | "SQL interpretation" |
| Standard | witness basis ($\alpha_w$) [2] | propagation ($\alpha_p$) [3] |
| QRI | minimal witness basis ($\alpha_w^m$) [2] | default-all propagation ($\alpha_p^d$) [1] |

**Figure 1: Particular definitions (naive, standard, QRI) for why- and where-provenance considered in this paper.**

Due to space constraints and in order to keep this paper to the point, we will assume basic familiarity of the reader with the provenance concepts given in Fig. 1 and not repeat their formal definitions. Instead, we refer to the detailed survey of Cheney et al. [4] from which we also borrow the running example of Fig. 2 and Fig. 4 (and the milk example after giving a real-world interpretation to the annotations). Appendix A summarizes the notation used throughout this paper.

## 2 The minimal witness basis as QRI why-provenance

Why-Provenance identifies witness tuples: "What input tuples contribute to the presence of each output tuple?" A *witness* is subset of the input tuples that is sufficient to ensure that a given output tuple $t$ appears in the result of a query. This definition implies that the whole database is a witness as it is sufficient for $t$ to appear. The *witness basis* or why-provenance $\alpha_w(t,Q)$ is a subset of only relevant witnesses where the definition by Buneman et al. [2] makes precise what "relevant" means. Intuitively, those tuples that have been involved in some operation during query evaluation are part of the witness basis. It turns out that why-provenance is not QRI, and relationally equivalent queries may have different witness bases.

Buneman et al. [2] showed that a subset of the witness basis, called the *minimal witness basis* and written here as $\alpha_w^m(t,Q)$, is invariant under rewriting. It con-

**(a)** $R$

| | $A$ | $B$ |
|---|---|---|
| $t_1$ | 1 | 2 |
| $t_2$ | 1 | 3 |
| $t_3$ | 2 | 2 |

**(b)** $Q_1, \alpha_w(Q_1)$

| | $A$ | $B$ | |
|---|---|---|---|
| $t_4$ | 1 | 2 | $\{\{t_1\}\}$ |
| $t_5$ | 1 | 3 | $\{\{t_2\}\}$ |
| $t_6$ | 2 | 2 | $\{\{t_3\}\}$ |

**(c)** $Q_2, \alpha_w(Q_2)$ **(d)** $\alpha_w^m(Q_2)$ **(e)** $\alpha_l(Q_2)$

| | $A$ | $B$ | (c) $\alpha_w(Q_2)$ | (d) $\alpha_w^m(Q_2)$ | (e) $\alpha_l(Q_2)$ |
|---|---|---|---|---|---|
| $t_4$ | 1 | 2 | $\{\{t_1\},\{t_1,t_2\}\}$ | $\{\{t_1\}\}$ | $\{t_1,t_2\}$ |
| $t_5$ | 1 | 3 | $\{\{t_2\},\{t_1,t_2\}\}$ | $\{\{t_2\}\}$ | $\{t_1,t_2\}$ |
| $t_6$ | 2 | 2 | $\{\{t_3\}\}$ | $\{\{t_3\}\}$ | $\{t_3\}$ |

**Figure 2: (a): Input table $R$. (b,c): Identical queries $Q_1(x,y) :- R(x,y)$ and $Q_2(x,y) :- R(x,y), R(x,_)$ together with the why-provenance $\alpha_w$ of their tuples. (d,e): Lineage $\alpha_l$ and minimal witness basis $\alpha_w^m$ of the tuples for $Q_2$.**

sists of all the *minimal* witnesses in the witness basis, where a witness is minimal if none of its proper subsets is also a witness. For example, the why-provenance of $t_4$ in $Q_2$ in Fig. 2c is $\alpha_w(t_4,Q_2) = \{\{t_1\},\{t_1,t_2\}\}$, however $\alpha_w^m(t_4,Q_4) = \{\{t_1\}\}$ in Fig. 2d since $\{t_1\} \subset \{t_1,t_2\}$, and thus, $\{t_1,t_2\}$ is not minimal.

Lineage $\alpha_l(t,Q)$ for an output tuple $t$ is a subset of the input tuples which are relevant to the output tuple, where the definition by Cui and Widom [5] makes precise what "relevant" means. Intuitively, we can get the lineage by taking the union over all witnesses in the why-provenance. We write this as $\alpha_l(t,Q) = \Cup \alpha_w(t,Q)$. For example, $\alpha_l(t_4,Q_2) = \Cup \alpha_w(t_4,Q_2) = \Cup\{\{t_1\},\{t_1,t_2\}\} = \{t_1,t_2\}$ in Fig. 2e.

## 3 Default-all propagation as QRI where-provenance

Where-provenance focuses on cells $(t,A)$, i.e. tuples $t$ together with an attribute $A$, and identifies witness cells: "Where (from what cell) does an output tuple value come from?" Hence, where-provenance of a cell $(t,A)$ consists of locations or values that can be found in tuples of the why-provenance of $t$. Since where-provenance was investigated in the context of propagating annotations from input to output cells [3], we write $\alpha_p(t,A,Q)$ for where-provenance (cp. Fig. 1).

Where-provenance is also not the same for equivalent queries, and there are two distinct issues to consider: (1) The first has to do with the way we write a conjunctive query in SQL (thus called "SQL interpretation" in Fig. 1), and is illustrated with Fig. 3: Query $Q_3'$ selects attribute $A$ from table $R$, whereas $Q_3''$ selects it from table $S$. Hence, a naive interpretation of propagation through SQL would lead to propagated values $\alpha_p^*(t_8,Q_3') = \{a,c\}$ versus $\alpha_p^*(t_8,Q_3'') = \{g\}$. This problem disappears once we consider Datalog notation, and got taken care of by the definition of propagation rules in [3] which propagate annotations from attributes of both joined tables.

(2) Secondly, certain relational rewrites do not preserve annotation propagation. Figure 4 gives a detailed example taken from [4] that shows that relationally equivalent queries $Q_1$ and $Q_2$ result in different annota-

**(a)** $R^a$

| | $A$ | $B$ |
|---|---|---|
| $t_1$ | $1^a$ | $2^b$ |
| $t_2$ | $1^c$ | $3^d$ |
| $t_3$ | $2^e$ | $2^f$ |

**(b)** $S^a$

| | $A$ | $C$ |
|---|---|---|
| $t_7$ | $1^g$ | $2^h$ |

**(c)** $Q_3, \alpha_P(Q_3)$

| | $A$ | $B$ |
|---|---|---|
| $t_8$ | $1^{a,c,g}$ | $2^h$ |

**(d)** $Q_3'$

```
SELECT  distinct R.A, S.C
FROM    R,S
WHERE   S.C = 2
```

**(e)** $Q_3''$

```
SELECT  distinct S.A, S.C
FROM    R,S
WHERE   S.C = 2
```

**Figure 3: A naive "SQL interpretation" of query $Q_3(x,y) :- R^a(x,y), S^a(x,\text{'2'})$ would lead to different where-provenances for cell $(t_8,A)$ in the output depending on whether SQL queries $Q_3'$ or $Q_3''$ were used.**

tions of their output (cp. Fig. 4b vs. Fig. 4c).

In an attempt to define a QRI propagation scheme for where-provenance, Bhagwat et al. [1] define the *default-all propagation* scheme, written here as $\alpha_p^d(t,A,Q)$. Their system DBNotes achieves QRI for where-provenance by including the provenance of all relationally equivalent rewrites $Q'$ for a query $Q$:

$$\alpha_p^d(t,A,Q) := \bigcup_{Q' \equiv Q} \alpha_p(t,A,Q')$$

For example, Fig. 4d shows the result annotations for both equivalent queries $Q_1$ and $Q_2$ over the input table Fig. 4a in the default-all propagation scheme. Intuitively, for both $Q_1$ and $Q_2$, default-all propagation returns the where-provenance of the relationally equivalent query $Q(x,y) :- R^a(x,y), R^a(_,y), R^a(x,_)$.

## 4 Default-all propagation is dangerous!

We next illustrate that the QRI property of default-all comes at a high price, namely the price of propagating *irrelevant* tuples to the output. This can be dangerous.

EXAMPLE 1 (MILK). *Hanako lives in Tokyo and worries about the recent nuclear accidents at the Fukushima nuclear power plant. She likes to drink*

**(a)** $R^a$

| | $A$ | $B$ |
|---|---|---|
| $t_1$ | $1^a$ | $2^b$ |
| $t_2$ | $1^c$ | $3^d$ |
| $t_3$ | $2^e$ | $2^f$ |

**(b)** $Q_1, \alpha_p(Q_1)$

| | $A$ | $B$ |
|---|---|---|
| $t_4$ | $1^a$ | $2^b$ |
| $t_5$ | $1^c$ | $3^d$ |
| $t_6$ | $2^e$ | $2^f$ |

**(c)** $Q_2, \alpha_p(Q_2)$

| | $A$ | $B$ |
|---|---|---|
| $t_4$ | $1^{a,c}$ | $2^b$ |
| $t_5$ | $1^{a,c}$ | $3^d$ |
| $t_6$ | $2^e$ | $2^f$ |

**(d)** $Q_2, \alpha_p^d(Q_2)$

| | $A$ | $B$ |
|---|---|---|
| $t_4$ | $1^{a,c}$ | $2^{b,f}$ |
| $t_5$ | $1^{a,c}$ | $3^d$ |
| $t_6$ | $2^e$ | $2^{b,f}$ |

**(e)** $Q_2, \alpha_p^m(Q_2)$

| | $A$ | $B$ |
|---|---|---|
| $t_4$ | $1^a$ | $2^b$ |
| $t_5$ | $1^c$ | $3^d$ |
| $t_6$ | $2^e$ | $2^f$ |

**Figure 4: (a): Annotated table $R^a$. (b,c): Equivalent queries $Q_1(x,y) :- R^a(x,y)$ and $Q_2(x,y) :- R^a(x,y), R^a(x,_)$ with the *where-provenance* $\alpha_p$ of their cells. (d,e): QRI variants *default-all propagation* $\alpha_p^d$ and *minimal propagation* $\alpha_p^m$.**

**(a)** $R^a$

| | *Food* | *Content* |
|---|---|---|
| $t_1$ | LF Milk | Cesium-137$^b$ |
| $t_2$ | LF Milk | Calcium$^d$ |
| $t_3$ | SC Water | Cesium-137$^f$ |

**(b)** $Q_4, \alpha_p^d(Q_4)$

| | *Content* |
|---|---|
| $t_9$ | Cesium-137$^{b,f}$ |

**(c) Annotation** $b$

| |
|---|
| user: Bob<br>date: March 18, 2011, 8:43pm<br>I have just measured half a glass of milk with my Geiger counter. I found five times the allowed amounts of Iodine-131 and Cesium-137. I will make a second measurement tomorrow to confirm. |

**(d) Annotation** $f$

| |
|---|
| user: Fuyumi<br>date: March 19, 2011, 7:25am<br>I measured 250ml bought yesterday and today, and I can assure you I found only small, negligible traces. |

**Figure 5: (a): Database for Example 1. Note that table $R^a$ is semantically the same as $R^a$ in Fig. 4a taken from [4]. (b): The query is $Q_4(y) :- R^a('\text{LF Milk}', y)$, i.e. "find all annotations for LF Milk." (c,d): Content of annotations $b$ and $f$.**

*lactose-free milk, but has just heard that traces of radioactive Cesium-137 were found in LF Milk of the local store. She is worried, and not so without reason. She queries a community database (Fig. 5a) for the content of LF Milk. The database includes data and user-generated annotations. She wants to make sure that she gets all relevant information and therefore opts for the default-all propagation scheme of user-generated community annotations (she is not familiar with databases and provenance, but "default-all" just sounds like the right thing to do). The database returns Fig. 5b with two annotations: b and f shown in Fig. 5c and Fig. 5d.*

*Based on the annotations the database returns, she decides to buy and drink the milk. Fuyumi is a very reputable friend of hers, and Fuyumi claims in the most recent annotation f that her measurements shows only low levels of radiation. However, what Hanako does not realize (and what the database does not expose to her) is that Fuyumi's comment has nothing to do with LF Milk. The comment propagated to the output because the database included annotations from all relationally equivalent queries. One such query is* $Q_4'(y) :- R^a('LF\ Milk', y), R^a(_, y)$*, which is responsible for propagating to the output an annotation about Cesium-137 in SC Water, a completely different product.*

*Basically, the default-all propagation scheme has given Hanako semantically irrelevant annotations, based on which she then made the wrong decision.*

# 5 Non-dangerous QRI where-provenance

Why is default-all propagation dangerous? The reason is a *mismatch in the semantics*. Just because two different tuples have the same value in an attribute does not imply that the annotations of those attribute values are related in any way. And, whereas rewriting the query $Q_1$ into query $Q_2$ with an additional (and unnecessary) self-join on table $R$ does not change the output tuples, we now have a join with *semantically irrelevant* tuples that propagates *irrelevant* information. And since the first step of making the scheme QRI (that of avoiding the issue in Fig. 3) propagates annotations from all cells that join, default-all propagation will make sure that completely irrelevant annotations propagate to the query output.

We propose instead the *minimal propagation* scheme. Intuitively, for a given output cell $(t,A)$, we intersect the where-provenance $\alpha_p(t,A,Q)$ with the annotations in the minimal witness basis $\alpha_w^m(t,Q)$ on all attributes $A'$ contributing to $\alpha_p$. Written differently:

$$\alpha_p^m(t,A,Q) := \bigcup_{\substack{t' \in \Uplus \alpha_w^m(t,Q) \\ A' \in \text{attributes of } t' \text{ propagating to cell}(t,A)}} \alpha_p\big(t', A', R'(t')\big)$$

Here, the expression $\Uplus \alpha_w^m(t,Q)$ transforms the minimal witness basis as sets of sets of tuples into a set of tuples (hence, it can be interpreted as a form of QRI lineage). The overall expression unions from all tuples $t'$ in the minimal witness basis, the annotations $\alpha_p$ from all attributes $A'$ of input table $R'(t')$ from which tuple $t'$ propagated values to the cell $(t,A)$. Since those attributes

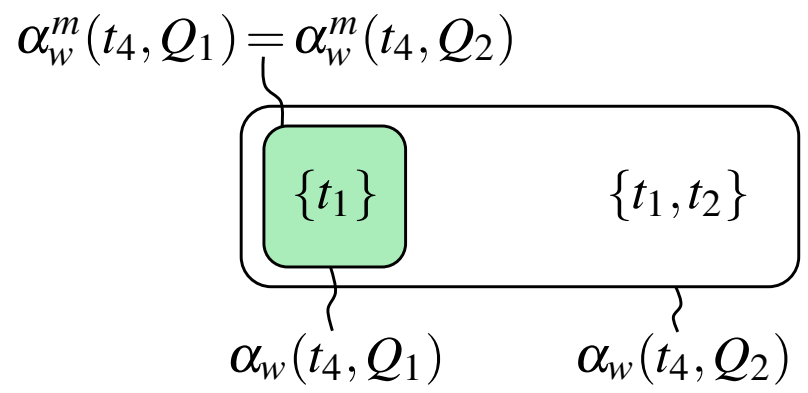

**(a) QRI why-provenance for tuple $t_4$ in Fig. 2: *minimal witness basis* $\alpha_w^m$ in green.**

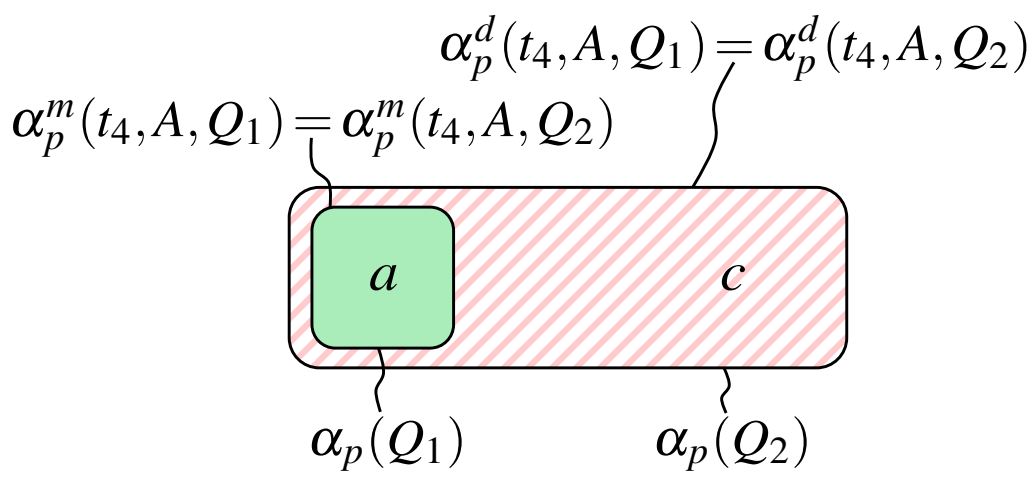

**(b) QRI where-provenance for cell $(t_4,A)$ in Fig. 4: *minimal propagation* $\alpha_p^m$ in green vs. *default-all* $\alpha_p^d$ in red.**

**Figure 6: (a) The *minimal witness basis* $\alpha_w^m$ considers only minimal sets of witnesses that imply the output. (b) In contrast, *default-all propagation* $\alpha_p^d$ considers the union of annotations for all equivalent queries. We propose instead *minimal propagation* $\alpha_p^m$ as QRI where-provenance analogous to the minimal witness basis $\alpha_w^m$ for why-provenance.**

are never changed by rewriting a query into an equivalent query, the output is well-defined and QRI.

The minimal propagation scheme has the following desirable properties:

(i) Just like default-all propagation, it is also QRI.
(ii) There is no need to evaluate any rewrite of a given query.[1]
(iii) Just as the minimal witness basis for why-provenance, it considers a minimal and QRI set of values (see Fig. 6). The intuition is that, among all relationally equivalent queries, those that have *no irrelevant self-join* (cp. $Q_1$ vs. $Q_2$ in Fig. 4) are the ones that most closely capture the user's actually intended semantics of the query.

For our running example, Fig. 4d and Fig. 4e compare the output of default-all with that of minimal propagation. For example, both where-provenance and default-all propagation return $\{a,c\}$ for output cell $(t_4,A)$ in query $Q_2$. In contrast, minimal propagation is $\{a\}$, because $t_1$ from $R^a$ is the only tuple in the minimal witness basis ($\Uplus\alpha_w^m(t_4,Q_2) = \{t_1\}$) with one contributing attribute $A$. Hence, $\alpha_p^m(t_4,A,Q_2) = \alpha_p(t_1,A,R^a)$ = {a}.

In our milk example (Example 1), minimal propagation gives the *only relevant* annotation $b$.

[1] Bhagwat et al. [1] provide an optimization that avoids having to evaluate infinitely many equivalent formulations for default-all, and it suffices to evaluate only a finite number.

# 6 Conclusions

Arguably, the QRI (query-rewrite-insensitive) property of annotation propagation is desirable. We do not discuss here whether this is indeed the case, but merely point out that, *if aiming for QRI*, care has to be taken not to trade a meaningful semantics in exchange for this property.

We illustrated that the *default-all* propagation scheme achieves QRI by including annotations from relationally equivalent but somehow irrelevant rewrites. This can lead to spurious annotations in the output which are *semantically irrelevant*, and thus can give the user a wrong impression of relevance. Hence, default-all is dangerous.

We proposed *minimal* propagation which is QRI, has a clean and simple semantics, and propagates all relevant and only relevant annotations to the output.

**Acknowledgements.** This work was partially supported by NSF grants IIS-0915054 and IIS-0911036, and was performed in the context of the Causality in Database project (`http://db.cs.washington.edu/causality/`).

# A Notation

| | |
|---|---|
| $t_i$ | input or output tuple |
| $R,S$ | input tables \| sets of tuples |
| $A,B,C$ | attributes of a table |
| $Q_i$ | queries or output tables |
| $\alpha_w(t,Q)$ | why-provenance (witness basis) for tuple $t$ of table $Q$ \| if context is known, also used as $\alpha_w(Q)$ or $\alpha_w(t)$ \| set of sets of tuples |
| $\alpha_w^m()$ | minimal witness basis |
| $\alpha_l(t,Q)$ | lineage of tuple $t$ in table $Q$ \| set of tuples |
| $R^a, S^a$ | annotated input tables |
| $\alpha_p(t,A,Q)$ | where-provenance (propagation) for the value of cell $(t,A)$ of table $Q$ \| if context is known, also used as $\alpha_p(t,A)$ or $\alpha_p(Q^a)$ \| set of values |
| $\alpha_p^d()$ | default-all propagation |
| $\alpha_p^m()$ | minimal propagation |